\documentclass[prd,a4paper,showpacs,nofootinbib,preprintnumbers,amsmath,amssymb, multicol]{revtex4}

\usepackage{epsfig}
\usepackage{amssymb}
\usepackage{graphicx}

\usepackage{color}
\definecolor{rosso}{cmyk}{0,1,1,0.4}
\definecolor{rossos}{cmyk}{0,1,1,0.55}
\definecolor{rossoc}{cmyk}{0,1,1,0.2}
\definecolor{blu}{cmyk}{1,1,0,0.3}
\definecolor{blus}{cmyk}{1,1,0,0.6}
\definecolor{bluc}{cmyk}{1,1,0,0.1}
\definecolor{verde}{cmyk}{0.92,0,0.59,0.25}
\definecolor{verdec}{cmyk}{0.92,0,0.59,0.15}
\definecolor{verdes}{cmyk}{0.92,0,0.59,0.4}

\def\BRed  {}
\def\Black{}
\def\Orange{} % PANTONE 177
\def\Green{} % PANTONE 323
\def\Blue {}
\def\circa#1{\,\raise.3ex\hbox{$#1$\kern-.75em\lower1ex\hbox{$\sim$}}\,}

\makeatletter
\def\art{\@ifnextchar[{\eart}{\oart}}
\def\eart[#1]#2#3#4#5#6{{\rm #2}, {\em #3 \rm #4} {\rm (#6) #5 ({\em #1})}}
\def\hepart[#1]#2{{\rm #2, \em#1}}
\newcommand{\oart}[5]{{\rm #1}, {\em #2 \rm #3} {\rm (#5) #4}}

\newcommand{\beq}{\begin{equation}}
\newcommand{\eeq}{\end{equation}}
\newcommand{\bea}{\begin{eqnarray}}
\newcommand{\eea}{\end{eqnarray}}
\newcommand{\ba}{\begin{array}}
\newcommand{\ea}{\end{array}}
\newcommand{\bi}{\begin{itemize}}
\newcommand{\ei}{\end{itemize}}
\newcommand{\bn}{\begin{enumerate}}
\newcommand{\en}{\end{enumerate}}
\newcommand{\bc}{\begin{center}}
\newcommand{\ec}{\end{center}}

\newcommand{\gsim}{\lower.7ex\hbox{$\;\stackrel{\textstyle>}{\sim}\;$}}
\newcommand{\lsim}{\lower.7ex\hbox{$\;\stackrel{\textstyle<}{\sim}\;$}}

\begin{document}
%\tolerance=100000
%\thispagestyle{empty}
\setcounter{page}{0}

%\preprint{XXXXXXXXXXX}

\title{The Cold Spot as a Large Void: \\  Rees-Sciama effect on CMB Power Spectrum and Bispectrum}

\author{Isabella Masina}  \email{isabella.masina@cern.ch}
\author{Alessio Notari}   \email{alessio.notari@cern.ch}

\affiliation{  {\it  CERN, Theory Division, CH-1211 Geneva 23, Switzerland}}

%\date{\today}

\begin{abstract}
~~~\\
\begin{center} {\bf Abstract} \end{center}
 
The detection of a ``Cold Spot" in the CMB sky could be explained by the presence of an anomalously large 
spherical underdense region (with radius of a few hundreds ${\rm Mpc}/h$) located between us and the 
Last Scattering Surface.
Modeling such an underdensity with an LTB metric, we investigate whether 
it could produce significant signals on the CMB power spectrum and bispectrum, via the Rees-Sciama effect. 
We find that this leads to a bump on the power spectrum, that corresponds to an ${\cal O}(5\%-25\%)$ correction at multipoles 
$5 \leq \ell \leq 50$; 
in the cosmological fits, this would modify the $\chi^2$ by an amount of order unity.
We also find that the signal should be visible in the bispectrum coefficients with a signal-to-noise $S/N\simeq {\cal O} (1-10)$, 
localized at $10 \leq \ell \leq 40$. Such a signal would lead to an overestimation of the primordial $f_{NL}$ 
by an amount $\Delta f_{NL}\simeq 1$  for WMAP and by $\Delta f_{NL}\simeq 0.1$ for Planck. 

\end{abstract}

% Particle and field theory models of early universe, Observational Cosmology, Voids, Background Radiation

\setcounter{page}{1}
%\tableofcontents

%\newpage

\maketitle

%%%%%%%%%%%%%%%%%%%%%%%%%%%%%%%%%%%%%%%%%%%%%%%%%%%%%%%%%%%%%%%%%%%%%%%%%%%%%%%%%%%%%%%%
\section{Introduction}

The recent WMAP~\cite{WMAP} experiment has measured with great accuracy the anisotropies of the Cosmic Microwave Background (CMB), 
whose features are in good agreement with the expectations from inflation of a Gaussian spectrum of adiabatic fluctuations, 
fully described by a nearly scale-invariant power spectrum.
However, it has been pointed out by several authors that the data contain various unexpected features.
Some of them are localized at very large angular scales, such as the low quadrupole, the alignment of the low multipoles 
(for a review see~\cite{reviewlowl} and references therein) and the power asymmetry between the northern and southern 
hemispheres~\cite{hemispherical}. 
Another anomaly is the presence of the so-called Cold Spot~\cite{ColdSpot1,ColdSpot2,ColdSpot3}, which is a large circular region 
on an angular scale of about $10^\circ$ that appears to be anomalously cold: the probability that such a pattern would 
appear from Gaussian primordial fluctuations is estimated to be about $1.8\%$~\cite{ColdSpot1,ColdSpot2,ColdSpot3}. 
So, while this could still be a statistical fluke, some authors have put forward the idea that it could instead be due
to an anomalously large spherical underdense region
of some unknown origin, on the line of sight between us and the Last Scattering Surface (LSS)~\cite{Tomita,InoueSilk}. 
We may also remind that~\cite{rudnick} has claimed that, looking at the direction of the Cold Spot in the Extragalactic 
Radio Sources (the NVSS survey), an underdense region is visible at redshift $z\sim 1$ (see, however~\cite{huterer} for a paper 
that challenges this claim). Another motivation for studying such objects is that an underdense region of two or three hundreds of ${\rm Mpc}/h$ 
could be enough to give an acceptable fit to the Supernova data and the CMB without Dark Energy~\cite{ABNV}, 
if we happen to live near its centre.

In this paper we also take the point of view that the Cold Spot could be due to such an underdense region, customarily denoted as a ``Void''. 
By modeling it through an inhomogeneous Lema\^itre-Tolman-Bondi (LTB) metric (which also requires an overdense compensating shell),
we compute the impact of such a Void on some observational quantities, focusing on the statistical 
properties of the CMB: 
in particular on the two-point correlation function (power spectrum) and the three-point correlation function (the bispectrum).
This is interesting for the following reasons.
First, if there is such a Void, does the power spectrum get a sizable correction? 
And, if yes, to what extent the estimation of the cosmological parameters is affected?
Second, if there is such a Void, could it be detectable also in the bispectrum?
Third, does the presence of such an object interfere with the measurement of a primordial non-gaussianity, 
which constitutes a very important piece of information in order to distinguish between models of inflation?

%On the other hand the presence of such a Void along the line of sight has a more subtel effect. At first order in the amplitude 
%of density fluctuations the Newtonian potential is constant in such a structure: and therefore a photon that enters the potential well, 
%exits with the same energy.
%However this is not true at second order and the photons gets some addistional blueshift (the so-called Rees-Sciama effect).
It is well known that, passing through a Void, photons suffer some blue-shift due to the fact that the gravitational potential is not 
exactly constant in time -- the so-called Rees-Sciama (RS) effect~\cite{ReesSciama}. By RS effect we mean the one associated to 
the variation of the gravitational potential from non-linear effects. There can be an effect already at the linear level, 
usually referred to as the Integrated Sachs-Wolfe (ISW) effect, which however would vanish in a matter dominated flat Universe.
The ISW effect would be significant only when a Dark Energy component becomes dominant with respect to matter.
Hence, here we focus our attention only on the RS effect, which is always present, and briefly comment on the extension to 
a $\Lambda$CDM Universe later on. Note that there is a second physical effect due to a Void on the line of sight, in addition to 
the RS blue-shift of the photons: the lensing of the primordial perturbations, which will be analyzed in detail in a companion paper~\cite{LENS}.

It is also known that the RS effect scales as the third power of the comoving radius of the Void $L$ times the present value of the Hubble parameter 
$H_0$, $\Delta T/T \propto - (L H_0)^3$.
For Void sizes compatible with the expectations from the usual structure formation scenarios, 
the RS effect happens to be suppressed with respect to the primordial temperature fluctuations ~\cite{Fullana}.
In order to produce a signal comparable to the measured temperature anisotropy of $|\Delta T/T| \sim 10^{-5}$, the Void should be 
of very large size, {\it i.e.} $L \sim (200-300) {\rm Mpc}/h$: this would be at odds with the standard scenario 
of structure formation from Gaussian primordial fluctuations (at more than $10\sigma$~\cite{InoueSilk}). We may also mention that according to~\cite{manyvoids,Szapudi} there are many localized regions in the sky (both underdense and overdense) of about $100 {\rm Mpc}/h$, which would be responsible for the correlations between the CMB and the Large Scale Structures: as~\cite{SarkarVoids} has recently stressed the existence of these regions is already at odds with the usual structure formation scenario.   
Even though we do not address in this paper the issue of the primordial origin of such large objects, we mention some possibilities.
For example, one could consider certain models of inflation, such as the ones of the ``extended'' type~\cite{extended,extendedDMN,extendedBN}, with the possibility of tunneling events 
and nucleation of bubbles, which would appear as Voids in the sky today. Alternatively, one could also imagine 
the presence of non-Gaussian features in the primordial fluctuations which would seed a large Void 
or even non-conventional structure formation histories in the late-time Universe which could enhance the probability of having large Voids. 

In the case of an explanation of the Voids via tunneling events, it is interesting to note that if a nucleation process has small probability (per unit volume and time) 
compared to the Hubble rate (at some time during inflation) then the number of anomalous Voids could well 
be very small, such as having just one or a few of them in the present observable Universe. 
In addition, since the LSS is a rather thin shell whose volume is much smaller than the total volume inside it,
if there are only few Voids it is more likely that they are located along the line of sight, rather than at the LSS\footnote{It is possible, also, 
to imagine that some primordial process could generate a coherent spherical region on the LSS surface, with small density contrast, on top of the 
primordial Gaussian spectrum, and which would produce a $10^{-5}$ fluctuation. However, we do not consider here such possibilities.}.
This is an important point because a Void at the LSS would have a huge impact on the CMB and, as a consequence, its size 
would be strongly constrained: $L\lesssim 4 {\rm Mpc}/h$~\cite{emptyVoids,extendedDMN}. 
On the contrary, Voids on the line of sight are not subject to the latter constraint.

Note also that other authors have put forward alternative ideas, such as the idea 
that the Cold Spot could be due to a topological defect, in particular a cosmic texture~\cite{texture}. 
As we are going to discuss,
some of our considerations apply in a similar fashion if the origin of the Cold Spot is a texture, 
rather than a Void.

The paper is organized as follows. 
In sect.~\ref{profiles} we calculate the RS-induced temperature profile of the CMB photons passing through a large Void 
with the physical characteristics of the Cold Spot.
The impact on the power spectrum is discussed in sect.~\ref{powerspectrum}, while sect.~\ref{bispectrum} deals with the impact
on the bispectrum and on the overestimation of $f_{NL}$ that would be done by neglecting the presence of such a Void.    
Finally, in sect.~\ref{multivoids} we extend some of the considerations to the case of several objects in the sky. In fact, it is conceivable 
that if there is one such structure, there may be other ones, maybe of smaller size and less visible -- see {\it e.g}.~\cite{manyvoids, Szapudi} 
for some other candidates in the CMB sky.  
Our conclusions are drawn in sect.~\ref{concl}.

%%%%%%%%%%%%%%%%%%%%%%%%%%%%%%%%%%%%%%%%%%%%%%%%%%%%%%%%%%%%%%%%%%%%%%%%%%%%%%%%%%%%%

\section{A Void in the line of sight}
\label{profiles}

\setlength{\textheight}{22cm}

As anticipated in the introduction, we would like to consider the following cosmological configuration: an observer looking 
at the CMB through one spherical Void located at comoving distance $D$ from him. We assume that the object does not intersect the LSS 
and that the observer is not inside it.
The observer receives from the LSS the CMB photons, whose fluctuations we assume to be adiabatic, nearly scale-invariant and 
Gaussian (as those generated {\it e.g.} by the usual inflationary mechanism).
Given this configuration, the observer detects one particular realization of the primordial inflation-generated perturbations on the 
LSS {\it plus} the secondary effects due to this anomalous structure located in the line of sight. 
Our aim is to give the theoretical prediction for the two-point and three-point correlation functions, 
in order to compare them with the observations. 

As a first step we consider that the Universe and the LSS are just completely homogeneous, and study what kind of 
profile we should see in the sky, because of the existence of one inhomogeneous Void located between us and the LSS: 
we denote the temperature fluctuation obtained in this way as $\Delta T^{(RS)}/T$, where (RS) stands for Rees-Sciama.
As a second step, we consider that there are Primordial fluctuations $\Delta T^{(P)}/T$ present on the LSS
that will also be affected by the presence of the Void: in fact, they will be lensed by it, leading to an 
additional effect, $\Delta T^{(L)}/T$.
So, finally, the total temperature fluctuation will be given by:
\beq
\frac{\Delta T}{T}=
\frac{\Delta T}{T}^{(P)} +
\frac{\Delta T}{T}^{(RS)} +
\frac{\Delta T}{T}^{(L)} \, ,
\label{3temp}
\eeq
where by definition each $(i)$-labeled fluctuation is given by
\beq
\frac{\Delta T^{(i)}}{T}\equiv \frac{T^{(i)}-\bar{T}^{(i)}}{T}
\label{defT}
\eeq
with the bar representing the angular average over the sky and $T=\sum_i \bar{T}^{(i)}=2.73 K$.

At this point it is important to be more precise about the origin of the Void, relatively to the Primordial fluctuations. 
If such a structure comes from the same physical process as the Primordial Gaussian spectrum, then this would mean that there 
is some correlation between the two. However we will assume, from now on, that the location of the Void in the sky is {\it not} 
correlated at all with the Primordial temperature fluctuations coming from inflation.
This is true, for example, if such structures come from a different process, such as nucleation of bubbles.
Note that this is a conservative assumption since, in the presence of a correlation, the temperature correlation functions 
would be generically enhanced.

Then, we need to fix the properties of the Void in order to compute the RS (and Lensing) contributions. 
Physically, one may characterize such a Void via the following quantities: its comoving distance $D$ from us, 
its comoving radius $L$, its present-day density contrast at the centre $\delta_0$, and some shape 
for the density profile. 
As we will discuss, the RS fluctuation $\Delta T^{(RS)}/T$ turns out to be described by two parameters: 
its amplitude at the centre $A$ and its angular extension, characterized by the diameter $\sigma$ of the Cold region. 
We also need some shape for the temperature profile, which will be determined by choosing some shape for the density profile of the Void.
Let us discuss in more detail how the parameters $A$ and $\sigma$ are related to the physical parameters of the Void, namely $D, L, \delta_0$:
the amplitude $A$ mostly depends on the radius $L$ and the density contrast $\delta_0$, while it is is not very sensitive to $D$; 
the diameter $\sigma$ depends on the ratio $L/D$.
As explained in the next paragraph, we choose the values for $A$ and $\sigma$ phenomenologically. 
Clearly this leaves a degeneracy in the choice of the physical parameters, since we have 
three of them ($D$, $L$ and $\delta_0$) and only two observational constraints ($A$ and $\sigma$).
 
In order to fix a range for the amplitude $A$ we rely on the values given by~\cite{texture} as follows. 
The minimal temperature in real space observed in the Cold Spot area is $\Delta T\simeq -400 \mu K$~\cite{ColdSpot1}, but this value 
includes as well a contribution from the Primordial Gaussian profile, $\Delta T^{(P)}/T$. 
The authors of~\cite{ColdSpot1} use a temperature profile for the secondary effect, add the Gaussian fluctuation 
and fit the observational data: this leads to an estimation for the value of the secondary effect at the centre, 
given by $\Delta T=-(190\pm 80) \mu K$, or equivalently $A= (7 \pm 3) \times 10^{-5}$. 
In this paper we assume the latter range of values for $A$. 
This is probably not entirely accurate (the shape of the profile that we use is slightly different from theirs, since we consider a compensated 
Void, while the profile in~\cite{texture} corresponds to the profile due to a cosmic texture), but it should give a good estimate for 
the range of values of interest. In any case, our results can be easily rescaled for different values of $A$.
As for the angular size $\sigma$ of the profile, we proceed as follows. According to~\cite{ColdSpot2, ColdSpot3}, a shape that fits well 
the Cold Spot has a diameter of about $10^\circ$ for the very cold part, but we also show in all plots the extreme case $\sigma=18^\circ$: in fact as one can see in fig.1 of~\cite{ColdSpot3} the entire cold region extends up to such large size.

Having fixed the numerical values, we choose a metric to model such an inhomogeneous region. The choice that we make is a spherically 
symmetric LTB metric, which is matched to a homogeneous and isotropic Friedman-Lema\^itre-Robertson-Walker (FLRW) flat model. 
For simplicity we consider a flat FLRW Universe with $\Omega_M=1$. Including a dark energy component
would lead to an ISW effect already at the linear level, in addition to the RS effect that we consider, 
which is present in any case, even in absence of dark energy.
The effect of a cosmological constant in an analogous setup has been studied {\it e.g.} by~\cite{InoueSilk,SakaiInoue} and the two effects turn out to be similar, with the same dependence on the Void radius, but with some difference on the dependence on the density contrast $\delta_0$~\footnote{Specifically, 
the $\delta_0$ dependence is different in the presence of $\Lambda$: the second-order term (RS effect) is suppressed, but there 
is a non-zero linear term. For small density contrast, $\delta_0\ll 1$, the linear ISW would dominate in a $\Lambda$-dominated cosmology. 
The quadratic term dominates if $\delta_0$ is large or if $\Omega_M$ is close to 1. An interesting case is the one for 
intermediate values ($\delta_0\approx 0.3$): the two effects are roughly compensating and the calculation does not depend much on 
the value of $\Omega_M$.}.

Another important point is that our profile is a compensated Void, with the underdense central region surrounded by a thinner 
external overdense region. This is a requirement dictated by the matching conditions, which corresponds to the condition 
that the Void does not distort the outer FLRW metric. Physically, if the Void comes {\it e.g.} from a primordial bubble of true vacuum, 
this is a consistent requirement: in fact a bubble would have a thin wall with localized gradient energy, compensating 
for the lower energy contained in the true vacuum in the interior region.
Note that, because of this feature, the angular size of the entire LTB patch that we consider will be larger than $\sigma$, since 
we have a hot region as well. In any case, we show that the main contribution to the power-spectrum and bispectrum comes from the 
inner underdense region.

Given these specifications for $\Delta T^{(RS)}/T$ we now compute its shape in the following subsections.

%%%%%%%%%%%%%%%%%%%%%%%%%%%%%%%%%%%%%%%%%%%%%%%%%%%%%%%%%%%%%%%

\subsection{Temperature Profile}

\begin{figure}[t!]\begin{center}\begin{tabular}{c}
\includegraphics[width=7cm]{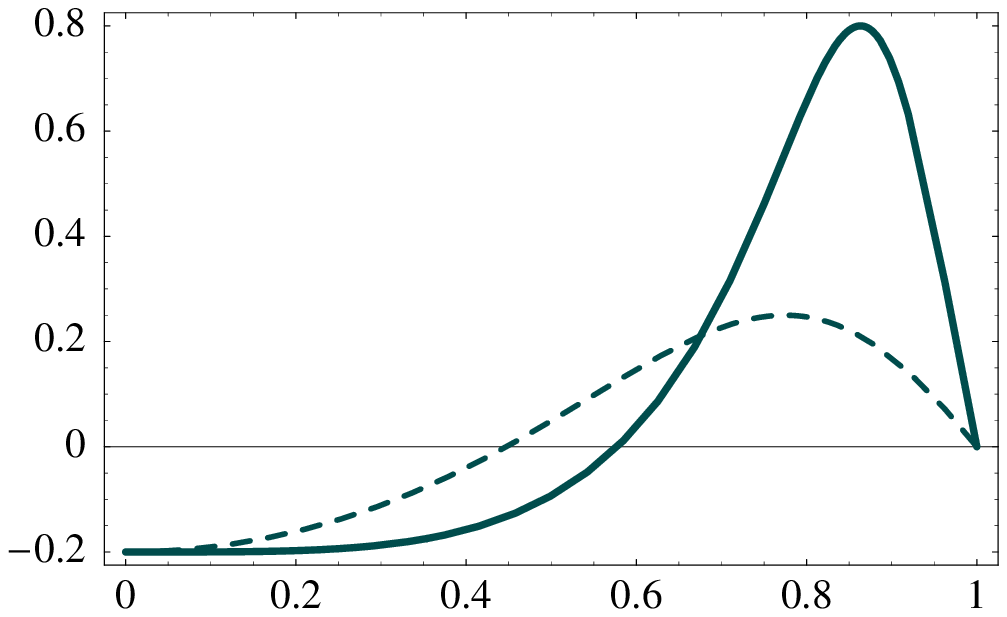} 
\put(-222,110){\Large $\delta(r)$} \put(5,-2){\Large $r/r_L$}
\put(-110,45){$\alpha=2$} \put(-67,100){$\alpha=4$}
\end{tabular}\end{center}\vspace*{-0.5cm} 
\caption{Plot of the present day density contrast $\delta(r)$, normalized for a central value of $\delta_0=-0.2$.
The solid (dashed) line corresponds to $\alpha=4$ ($\alpha=2$).}
\label{densita}
\vskip.5cm
\end{figure}

First of all, we model the profile as an LTB metric with irrotational dust, which describes a compensated Void.
It has been shown in~\cite{BN} that this metric can be treated, in some cases, as a perturbation of an FLRW metric 
with a gravitational 
potential $\Phi$ given by the following (Newtonian gauge) expression
\beq
ds^2=a^2(\tau) \left[ -(1+2 \Phi)d\tau^2+(1-2 \Phi) dx^i dx^j \right]    \, ,
\eeq
where $\tau$ is the conformal time and $x^i$ are dimensionless comoving coordinates. 
We have chosen units such that the present value of the conformal time is $\tau_0=(6\pi)^{-1/6}$.
As a function of the dimensionless comoving radial coordinate $r$, the gravitational potential is given by:
\beq
\Phi(r)=-\frac{9 \, 3^{1/3}}{5~ (2 \pi)^{2/3}} \int^r k(\bar{r}) \bar{r} d\bar{r}  \label{potenziale}    \, ,
\eeq
where $k(r)$ is an arbitrary function which represents the local curvature and determines the shape of the density profile. 
This approximation is valid as long as $k(r)$ is small. The only constraints on this function comes from the smoothness of the density profile at the centre 
(which dictates $k'(0)=0$, where the prime denotes a derivative with respect to the $r$ coordinate) and from the requirement that the LTB patch 
matches to a flat FLRW universe ($k'(L)=k(L)=0$).
We have chosen arbitrarily the function $k(r)$ as follows:
\beq
k(r)=k_0 \left[1-\left(\frac{r}{r_L}\right)^{\alpha} \right]^2~~~~~~~,~~~~~~r\leq r_L =  \frac{1}{2 (6 \pi)^{1/6}} L H_0 
\label{kappa}    \, ,
\eeq
where $\alpha>1$, $r_L$ is the dimensionless radius of the LTB patch and $H_0$ is the present value of the Hubble parameter. 
The density contrast that results from the choice of $k(r)$ is approximately given by the following expression~\cite{BN}:
\beq
\delta(r)\equiv\frac{\rho(r)-\bar{\rho}}{\bar{\rho}}=-\frac{B(r)}{1+B(r)} ~~~,~~~B(r)\equiv \frac{3^{4/3}}{10 (2\pi)^{2/3}}\tau^2 (k(r) r)'~,
\label{deltar}
\eeq
where $\rho(r)$ is the matter energy density inside the LTB patch, $\bar{\rho}$ is the average density in the outer FLRW region.
Note that the density contrast can be large even if $k(r)$ is small. We show in fig.~\ref{densita} the density profiles corresponding to certain values of $\alpha$. From~(\ref{kappa}) and~(\ref{deltar}) it follows that the value of the normalization 
constant $k_0$  is directly related to $\delta_0$ (the value of the present day density contrast at the centre of the Void):
$k_0=\frac{20 \pi \delta_0}{3(1-\delta_0)}$.

Given the potential $\Phi(r)$ we can compute the $\Delta T/T$ for a photon which travels through this patch, by computing 
a line integral, following the expression given in~\cite{MM,BMM} which is valid at second order in perturbation theory in $\Phi$:
\beq
\frac{\Delta T}{T}^{(RS)}= 2 \int^{\tau_E}_{\tau_O} d\tau \tau \left( \frac{1}{6}\Phi'^2-\frac{10}{21}\Upsilon_0  \right)   \, ,
\label{lineintegral}
\eeq
where $\tau_E$ is the value of the conformal time at emission (although of course the integrand is nonzero only in the region where the 
LTB structure is located). The conformal time evolves simply as $\tau(r)=(r-r_O)+\tau_O$, 
where the subscript $O$ refers to the observer space-time point.
Here $\Upsilon_0$ is a non-local quantity, which is given in terms of $\Phi$ as~\cite{BMM,BN}:
\beq
\Upsilon_0=-\int_{r_L}^r \frac{d\bar{r}}{\bar{r}^2} \int_{r_L}^{\bar{r}} d\tilde{r} 
\left[ \Phi(\tilde{r})'^2+2 \,\tilde{r} \,\Phi'(\tilde{r}) \Phi''(\tilde{r}) \right]  \, .
\eeq

\begin{figure}[t!]\begin{center}\begin{tabular}{c}
\includegraphics[width=8cm]{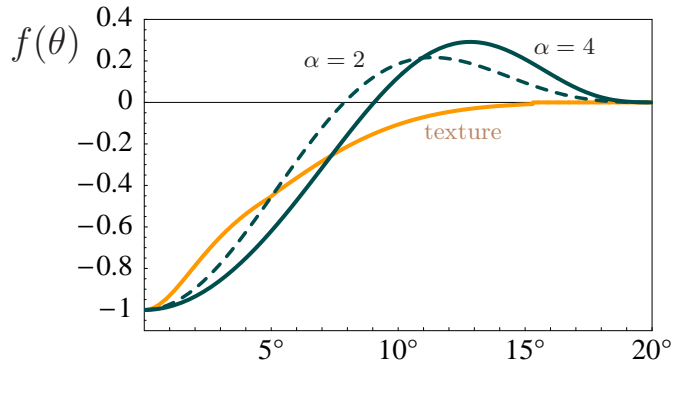} 
\put(-250,120){\Large $f(\theta)$} \put(0,-10){\Large $\theta$}
\put(-140,115){$\alpha=2$} \put(-54,120){$\alpha=4$}
\put(-95,88){\small \Orange texture \Black}
\end{tabular}\end{center}\vspace*{-0.5cm} 
\caption{Plot of the profile $f(\theta)$ for $\theta_L=20^\circ$.
The dark solid (dashed) line corresponds to $\alpha=4$ ($\alpha=2$) and a Cold Spot with diameter $\sigma=18^\circ$ ($\sigma=16^\circ$).
The light solid line shows, for comparison, the temperature profile (normalized so that $\Delta T/T=-1$ at the centre) 
of the texture that gives the best fit to the Cold Spot, as claimed in~\cite{texture}. }
\label{fig-p}
\vskip.5cm
\end{figure}

Then, using the spherical coordinates angles, $0\le\theta\le\pi$ and $0\le\phi < 2\pi$, we may parameterize the temperature profile as
\beq
\frac{\Delta T}{T}^{(RS)}(\theta,\phi)= \begin{cases}  A~ f(\theta) & {\rm if}~\theta < \theta_L \\ ~~0 & {\rm if}~\theta \ge \theta_L  \end{cases} ~~~~,
~~\tan \theta_L \equiv \frac{L}{D}~~,
\eeq
where the profile function $f(\theta)$ is normalized so that $f(0)=-1$. 
The profile has no dependence on $\phi$ because we choose the $\hat{z}$ axis to point towards the centre of the Void. 
The $A$ factor can be computed performing the integral~(\ref{lineintegral}) analytically along a radial trajectory, 
and it is a function of the physical parameters of the configuration: the radius $L$ of the LTB patch, the density contrast 
at the centre of the Void $\delta_0$ and the distance $D$ between the observer and the centre of the Void. 
The result is\footnote{The dependence on the shape of the density profile (the $\alpha$ parameter) comes in the numerical 
factor and it is given by $\frac{104 \alpha ^4}{7 \left(8 \alpha ^4+50 \alpha ^3+105 \alpha ^2+90 \alpha+27\right)}$, which ranges 
between 0.05 and 1.85. This variability in the prefactor could account for the prefactors obtained by previous 
analyses \cite{InoueSilk,SakaiInoue}, which are of about $0.1$.}, 
at leading order in $L H_0$,
\beq
A = \frac{26624}{51205} (L H_0)^3 \delta_0^2 \left(1 - \frac{D H_0}{2}\right)~~~,~~~{\rm for}~ \alpha=4~~.
\eeq
Note that the dependence on the distance $D$ is a weak correction, unless the patch happens to be located at distances comparable 
to the horizon.
We can compute the function $f(\theta)$ by numerical integration of~(\ref{lineintegral}) along a non-radial trajectory 
(computed as an unperturbed straight line, since any deviation would lead only to higher order corrections).
In fig.~\ref{fig-p} we plot the profile $f(\theta)$ as a function of $\theta$ for two values of $\alpha$.
For the reader's convenience, we provide polynomial interpolations of the profiles:
\bea
f(\theta)&=& -1 +  6.663 x^2 - 5.954 x^4 - 17.258 x^6 + 33.959 x^8 - 19.361 x^{10} + 2.940 x^{12} ~~~~~,~~  {\rm for}~\alpha=4 \nonumber \\ 
& &\\
f(\theta)&=& -1 + 11.191 x^2 - 37.576 x^4 + 58.272 x^6 - 46.190 x^8 + 17.904 x^{10} - 2.601 x^{12} ~~,~~  {\rm for}~\alpha=2 \nonumber \, ,
\eea
where $x\equiv \theta/\theta_L$.
We have added also a light solid line in fig.~\ref{fig-p} which shows, for comparison, the temperature profile of the cosmic texture that provides the best 
fit to the Cold Spot, as discussed in~\cite{texture} 
(its profile has also been normalized so that $\Delta T/T=-1$ at the centre). 
Notice that the compensated Void has a hot ring in the profile, while the texture does not have it~\footnote{There is one {\it caveat}, however: in a $\Lambda$ dominated
cosmology and with small contrast $\delta_0$, the linear ISW can become the dominant effect; and it has been shown by~\cite{InoueSilk} that this case does not lead
to a hot ring.}.

%We define $\tau_0 = 1/(6 \pi)^{1/6}$ and
%consider the i-th local void characterized today by $L_i [Mpc/h]$, $\sigma_i$, $\delta_i$. 
%Then $L_{ci}=H_0 L_i \tau_0/2$, $l_i=L_i/\tan \sigma_i$, 
%$l_{ci}=H_0 l_i \tau_0/2$, $z_i =\tau_0^2/(\tau_0-l_{ci})^2 -1$.

%%%%%%%%%%%%%%%%%%%%%%%%%%%%%%%%%%%%%%%%%%%%%%%%%%%%%%%%%%%%%%%%%%%%%%%%%%%%%%%%%%%%%%%%%%%%

\subsection{Decomposition into spherical harmonics} 

The spherical harmonic decomposition of the $(i)$-th profile for the temperature anisotropy $\Delta T^{(i)}(\hat {\bf n})/T$
of~(\ref{3temp}) is defined as:
\beq
a^{(i)}_{\ell m} \equiv \int d \hat {\bf n}~ \frac{\Delta T^{(i)}(\hat {\bf n})}{T}~ Y^*_{\ell m}(\hat {\bf n})
%~~~ \qquad , \qquad b_{\ell m} \equiv\int d \hat {\bf n}~ \Theta(\hat{\bf n}) ~ Y^*_{\ell m}(\hat {\bf n})
~~.  
\label{almblm}
\eeq

\begin{figure}[t!]\vskip 1cm
\begin{center} \begin{tabular}{c}
\includegraphics[width=7cm]{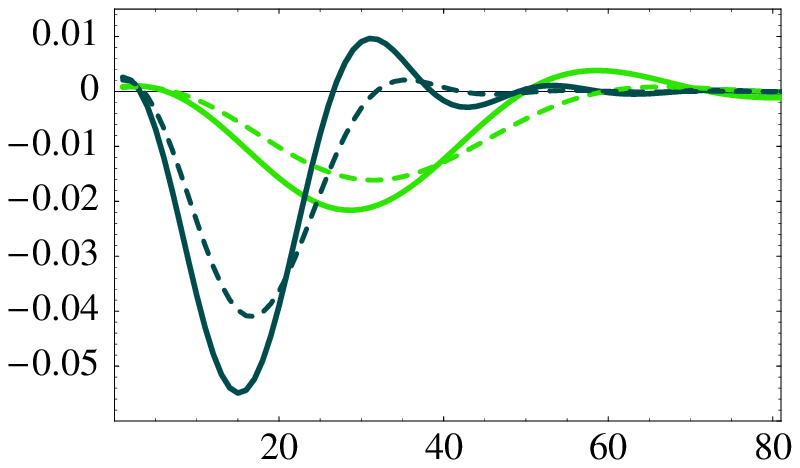} 
\put(-216,110){\Large $ \frac{a^{(RS)}_{\ell 0}}{A}$} \put(-0,0){\Large $\ell$}
\put(-130,28){\Blue $\theta_L=20^\circ$ \Black} \put(-97,65){\Green $\theta_L=11^\circ$ \Black}
\put(-55,30){solid: $\alpha=4$}
\put(-65,20){dashed: $\alpha=2$}
$~~~~~~~~~~~~~~$ 
\includegraphics[width=7cm]{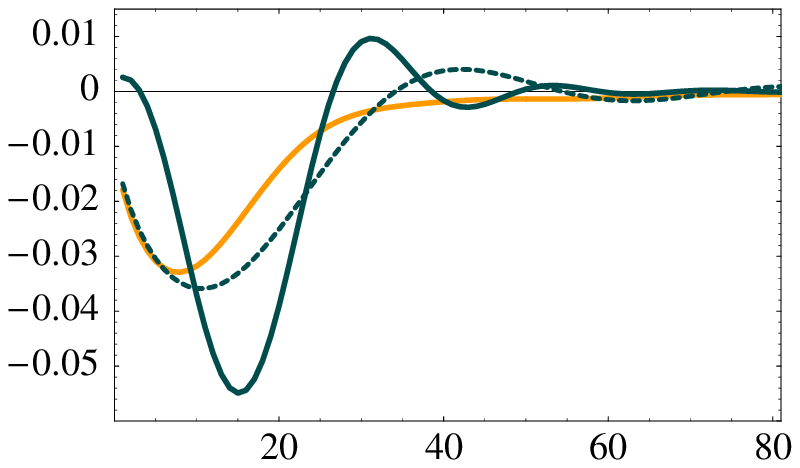} 
\put(-218,110){\Large $\frac{ a^{(RS)}_{\ell 0}}{A}$} \put(-0,0){\Large $\ell$}
%\put(-124,123){$\sigma=18^\circ$, $\alpha=4$}
\put(-131,29){\Blue $\nwarrow$ \Black}
\put(-127,21){\Blue compensated \Black}
\put(-113,79){\Blue $\nwarrow$  \Black}
\put(-112,72){\Blue non-compensated \Black}
\put(-150,88){\Orange \tiny texture \Black}
\put(-141,79){\Orange $\searrow$ \Black}
\end{tabular}\end{center}\vspace*{-0.5cm} 
\caption{Plots of $a^{(RS)}_{\ell 0}/A$ as a function of $\ell$. 
Left:  
The dark (light) solid line corresponds to a profile with $\alpha=4$ and $\theta_L=20^\circ (11^\circ)$, namely a Cold Spot 
with diameter $\sigma=18^\circ(10^\circ)$;
dashed lines correspond to similar profiles but with $\alpha=2$ and $\sigma=16^\circ (9^\circ)$.
Right: 
The solid line corresponds to the full compensated profile with $\alpha=4$ and $\theta_L=20^\circ$, while the dashed line 
to the non-compensated profile, namely the previous one truncated so to include only the cold region (Void). 
The light curve shows, for comparison, the ratio $a_{\ell 0}/A$ that one would obtain from the temperature profile of the  
texture giving the best fit to the Cold Spot~\cite{texture}.}
\label{almP} \vskip.5cm
\end{figure}

For the (RS) component, since the profile is symmetric with respect to the $\hat{z}$ axis pointing towards the centre of the Void, 
only the $a_{\ell m}$ coefficients with $m=0$ are non-vanishing (and they are real).
In the left plot of fig.~\ref{almP} we show the ratio $a^{(RS)}_{\ell 0}/A$ as a function of the multipole $\ell$. 
The dark solid (dashed) curve corresponds to a temperature profile with $\theta_L=20^\circ$ and $\alpha=4$ ($\alpha=2$),
hence to a Cold Spot with diameter $\sigma=18^\circ$ ($\sigma=16^\circ$); 
the light solid (dashed) curve is obtained by choosing instead $\theta_L=11^\circ$, so that $\sigma=10^\circ$ ($\sigma=9^\circ$). 
Notice that the medium amplitude of such a ratio is roughly equal to the fraction of the sky covered by our LTB patch,
namely about $3\%$ for $\theta_L=20^\circ$ and $1\%$ for $\theta_L=11^\circ$, respectively concentrated at multipoles
$10 \lesssim \ell \lesssim 20$ and $15 \lesssim \ell \lesssim 40$.

One may wonder whether the presence of a compensating hot shell~(see fig.~\ref{fig-p}) in our profile has a significant 
impact on the magnitude and shape of the $a^{(RS)}_{\ell 0}$ coefficients. As an example, in the right plot of fig.~\ref{almP} 
we focus on the $a^{(RS)}_{\ell 0}$ with $\theta_L=20^\circ$ and $\alpha=4$ (solid line), and show how they would change by truncating 
the temperature profile only to the cold part (dashed line). This approximately mimics a truncation of the density profile only to the 
underdense region (the true Void), disregarding the overdense compensating shell. 
One can see that the $a^{(RS)}_{\ell 0}$ are now suppressed but still quite similar, so that for our purposes the result is not going 
to be very different. This shows that the RS effect mainly comes from the underdense region. 
The same plot also shows, for comparison, the ratio $a_{\ell 0}/A$ that one would obtain from the temperature 
profile of the texture giving the best fit to the Cold Spot~\cite{texture}. Notice that the texture 
and the non-compensated Void are pretty similar.

%%%%%%%%%%%%%%%%%%%%%%%%%%%%%%%%%%%%%%%%%%%%%%%%%%%%%%%%%%%%%%%%%%%%%%%%%%%%%%%%%

\section{Power spectrum}
\label{powerspectrum}

Given a temperature profile with its $a_{\ell m}$ coefficients one can compute the associated two-point correlation function. 
In general, given {\it a single} set of $a_{\ell m}$ coefficients, the two-point correlation function (power spectrum) is defined 
via the $C_{\ell}$'s coefficients as
\beq
C_{\ell}\equiv\sum_{m=-\ell}^{\ell} \frac{|a_{\ell m}|^2}{2\ell+1}   \, .
\eeq
Note that this definition ensures that the $C_{\ell}$'s do not depend on the choice of the coordinate system on the sphere. 
Therefore we are free to keep our $\hat{z}$ axis aligned with the centre of the Void.
In our case $a_{\ell m}= a^{(P)}_{\ell m}+a^{(RS)}_{\ell m}+a^{(L)}_{\ell m}$ and, in order to face it with the experimentally 
observed value of $C_\ell$, we have to estimate the theoretical prediction for $\langle C_\ell \rangle$, where the $\langle...\rangle$ brackets 
stand for a statistical average over an {\it ensemble} of possible realizations of the Universe -- or, equivalently, an average over many 
distant uncorrelated observers.  

For a Primordial and Gaussian signal the two-point correlation functions are given by:
\beq
\langle a^{(P)}_{\ell_1 m_1} a^{(P) \, *}_{\ell_2 m_2} \rangle=\delta_{\ell_1 \ell_2} \delta_{m_1 m_2} \langle C^{(P)}_{\ell_1} \rangle   \ ,
\label{infpred}
\eeq
where the $\langle C^{(P)}_\ell \rangle$ are predicted by some mechanism ({\it i.e.} inflation) that can generate Primordial 
Gaussian fluctuations. 
Then, there are two types of effects on the power spectrum due to our secondary effects: \\
i) the inflationary prediction $C_\ell=\langle C^{(P)}_\ell \rangle$ receives corrections;\\ 
ii) there is also some non-diagonal correlation between different $\ell$'s.

As we have already stressed, we assume that the Rees-Sciama (RS) component in~(\ref{3temp}) is uncorrelated with 
the Primordial (P) component, which means that it can be factored out of the brackets. Given this fact, from~(\ref{3temp}) 
and~(\ref{almblm}), we get two types of non-zero contributions for $\langle C_\ell \rangle$: the RS-RS contribution proportional to
$|{a_{\ell m}^{(RS)}}|^2$, and the P-L contribution proportional to $\langle {a_{\ell m}^{(P)}}^* a_{\ell m}^{(L)} \rangle$.
The first effect is computed in the next subsection, while the second in the companion paper~\cite{LENS}.

%%%%%%%%%%%%%%%%%%%%%%%%%%%%%%%%%

\subsection{Rees-Sciama power spectrum}

\begin{figure}[t!] \begin{center}\begin{tabular}{c}
\includegraphics[width=7cm]{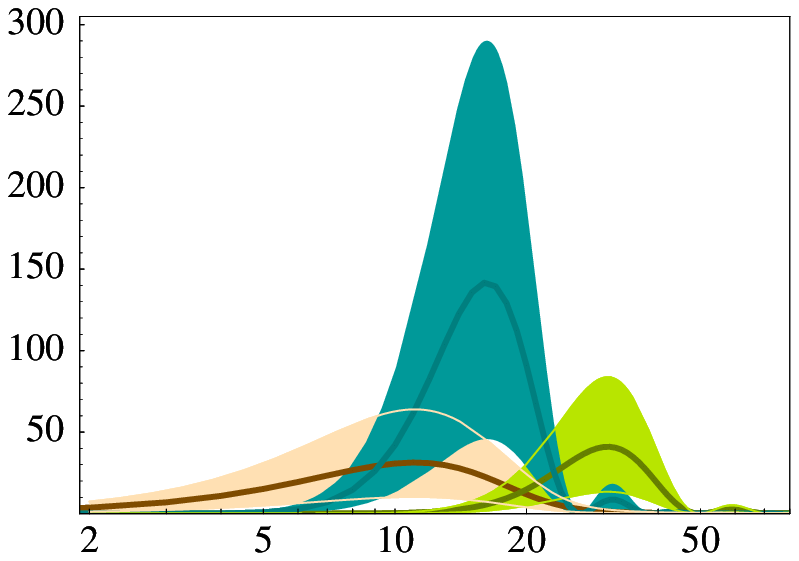} 
\put(-215,35){\rotatebox{90}{$ C^{(RS)}_{\ell} \frac{\ell (\ell+1)}{2 \pi} T_0^2 [\mu K^2]$}} \put(-0,0){\Large $\ell$}
\put(-65,94){\Blue $\sigma=18^\circ$ \Black} \put(-37,38){\Green $\sigma=10^\circ$ \Black}
\put(-149,25){\Orange \rotatebox{15}{\small texture} \Black}
~~~~~~~~~~~~~~~~~~ \includegraphics[width=7cm]{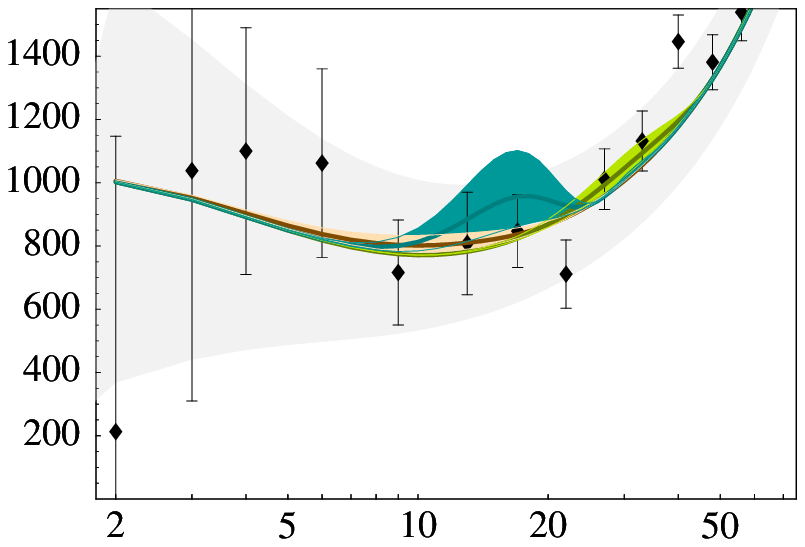} 
\put(-215,40){\rotatebox{90}{$ \langle C_{\ell}\rangle \frac{\ell (\ell+1)}{2 \pi} T_0^2 [\mu K^2]$}} \put(-0,0){\Large $\ell$}
\end{tabular}\end{center}\vspace*{-0.5cm} 
\caption{In the left panel we plot the $C_{\ell}^{(RS)}$'s from the Rees-Sciama effect.
The dark blue (light green) shaded region corresponds to an angular diameter for the Cold Spot equal to $\sigma=18^\circ (10^\circ)$, 
for a range of amplitudes $A = (7 \pm 3)\times 10^{-5}$.
For comparison, we also show (light orange) the analogous correction obtained in the case that the Cold Spot is due 
to a texture~\cite{texture}.
In the right panel the RS correction is added to the best-fit Primordial spectrum, from the usual 
$\Lambda$CDM concordance model; the cosmic variance (gray band) and the experimental binned data points are also shown.}
\label{fig-C2}\vskip 1cm
\end{figure}

As we have seen, for a spherical Void in the $\hat{z}$-direction, $a^{(RS)}_{\ell m}=0$ if $m\neq 0$.
Hence the RS contribution to the power spectrum coefficients, 
$\langle C_{\ell} \rangle = \langle C_{\ell}^{(P)} \rangle +  C_{\ell}^{(RS)}$, is just:
\beq
C^{(RS)}_{\ell}=\frac{|a_{\ell 0}^{(RS)}|^2}{2\ell+1}  \, .
\label{ClRS}
\eeq
In the left and right panels of fig.~\ref{fig-C2} we have plotted respectively the quantities 
$C^{(RS)}_{\ell} \frac{\ell (\ell+1)}{2 \pi} T_0^2$ and $\langle C_{\ell} \rangle \frac{\ell (\ell+1)}{2 \pi} T_0^2$, 
for the range of parameters mentioned in sect.~\ref{profiles}. The shaded regions are indeed obtained varying the amplitude
$A$ in its $1$-$\sigma$ range, namely $A= (7 \pm 3)\times 10^{-5}$. 
The results for different values of $A$ can be extracted just by noticing that $C^{(RS)}_{\ell} \propto A^2$.
As one can see, for a Void with a size which can account for the Cold Spot, the correction to the 
power spectrum is non-zero only in the range $5 \lesssim \ell \lesssim 50$ and its magnitude is about 
$5\%-25\%$ with respect to the Primordial signal.
We also note that the RS correction is of the order of the cosmic variance,  
$\Delta C_{\ell}^{P} = \langle C^{(P)}_{\ell} \rangle \sqrt{2/(2\ell+1)}$.
For comparison, we also show in the left panel of fig.~\ref{fig-C2} the correction to the power spectrum obtained from the texture of~\cite{texture},
which turns out to be smaller and shifted at lower multipoles with respect to the Void case.

In order to see how large is the impact on the cosmological parameter estimation due the RS effect from a large Void, 
one should perform a detailed statistical analysis of the CMB data, which we postpone for future work~\cite{HMN}.
For the purpose of this paper, we may just estimate the impact that the inclusion of the RS effect has on the $\chi^2$ of the fit,
which gets increased by the amount:
\beq
\Delta \chi^2=\sum_{2 \le \ell} \frac{{C_{\ell}^{(RS)}}^2}{\sigma^2_{\ell}}   \, ,
\eeq
where $\sigma^2_\ell$ is the variance of the two-point correlation function, including the cosmic variance and the instrumental noise.
For an experiment like WMAP, the cosmic variance is the dominant source of error at low-$\ell$'s, so that we can just neglect the 
instrumental noise and use $\sigma_{\ell}^2=\langle C^{(P)}_{\ell} \rangle^2 \frac{2}{2\ell+1}$.
The sum gives roughly the following result, for different angular diameters $\sigma$ of the Void:
\beq
\left(\frac{7\times 10^{-5}}{A}\right)^4 \Delta\chi^2 \simeq  
\begin{cases} 0.7 & {\rm for}~ \sigma=10^\circ \\ 3.6 & {\rm for}~ \sigma=18^\circ \end{cases} ~.
\eeq

Finally, as already mentioned, there is also a non-diagonal contribution to the two-point correlation function, 
which correlates different $\ell$'s in the range $5-50$ and is of the same order of magnitude as the 
diagonal contribution to the power spectrum, $C_\ell^{(RS)}$.

%The effect is not visible on a single $l$. When summing over all the $l$'s we can find a signal-to-noise ratio, defined as:
%\beq
%\frac{S}{N}\equiv \sum_l \frac{C^{RS}_l}{\sigma^2_{l}}=...
%\eeq
%We may actually include also the non-diagonal part in this signal-to-noise as
%\beq
%\Delta\chi^2|_{\rm non-diag}=\sum_{2<l_1<l_2} \frac{ a^{RS}_{l_1} a_{l_2}^{RS}}{\sigma^2_{l_1 l_2}}
%\eeq

%%%%%%%%%%%%%%%%%%%%%%%%%%%%%%%%%%%%%%%%%%%%%%%%%%%%%%%%%%%%%%%%%%%%%%%%%%%%%%%%%%

\section{Bispectrum}
\label{bispectrum}

Having computed the $a_{\ell m}$ coefficients and the two-point correlation functions, we now estimate the impact that a 
large Void has on the bispectrum coefficients, also to check whether this observable could be used to constrain the size
and the density of the Void itself, which is a highly non-Gaussian object. 
Moreover, since the bispectrum is the main tool used to make detections of a primordial non-gaussianity, it is 
interesting to see whether the presence of a large Void can contaminate this detection and to what extent.

Note that the bispectrum from the the RS effect due to conventional structures, {\it i.e.} with non-linearities at scales of order 
$10 \, {\rm Mpc}/h$, has already been studied by several authors~\cite{SpergelGoldberg, MatarreseRS}. However, the calculation 
here is quite different for two reasons: i) we deal with much larger values for the $a_{\ell m}$'s, because we consider larger values of $L$ and the 
RS effect on the temperature profile is proportional to $L^3$; 
ii) we are assuming no correlations between the RS and the Primordial fluctuations, 
while this is not the case in the conventional calculation where the RS effect arises from the structures seeded by the 
Primordial fluctuations themselves.
In addition, we compute in ~\cite{LENS} the associated Lensing effect.

The basic quantities are now the $B_{\ell_1~\ell_2~\ell_3}^{m_1 m_2 m_3}$ coefficients, defined as
\beq
B^{m_1 m_2 m_3}_{~\ell_1~ \ell_2 ~\ell_3} \equiv  ~ a_{\ell_1 m_1} ~ a_{\ell_2 m_2} ~ a_{\ell_3 m_3} \, ,
\label{defBlm}
\eeq
which are coordinate-dependent quantities. So, in analogy with the $C_\ell$'s coefficients, one introduces 
the angularly averaged bispectrum 
\beq
B_{\ell_1 \ell_2 \ell_3} = \sum_{m_1,m_2,m_3} 
\left( \begin{array}{ccc} \ell_1 & \ell_2 & \ell_3 \\ m_1 & m_2 & m_3  \end{array} \right)
%\left( \matrix{l_1 & l_2 & l_3 \cr m_1 & m_2 & m_3} \right) 
B^{m_1 m_2 m_3}_{~\ell_1 ~\ell_2 ~\ell_3}  \, ,
\label{defBl}
\eeq
where the matrix represents the Wigner 3-j symbols and the sum is carried over all possible values for the $m_i$'s.
One can indeed show, by using the properties of the Wigner 3-j symbols~\cite{review}, that this combination does not depend 
on the chosen $\hat{z}$-axis, so that these quantities are more suitable to make predictions. For convenience, we nevertheless
keep our $\hat{z}$-axis along the direction of the centre of the Void.

We are thus interested in evaluating $\langle B_{\ell_1 \ell_2 \ell_3} \rangle$ and, using~(\ref{3temp}) and~(\ref{defBl}), 
one realizes that it corresponds to a sum of 27 terms, of which 23 have zero statistical average. 
As already mentioned, a crucial assumption that we make here is that the coefficients $a^{(RS)}_{\ell 0}$ are {\it not} 
stochastic quantities, which means that the location of the Void in the sky is not correlated at all with the Primordial 
temperature fluctuations coming from inflation (this is a conservative assumption: if there is some correlation, 
the non-gaussianity could be much more important, since some terms would be non-zero). 
Under this assumption, the four types of terms that potentially survive are the ones involving: 
$\langle (a^{(RS)})^3 \rangle$, $\langle a^{(P)}a^{(L)} a^{(RS)} \rangle$, $\langle a^{(RS)} (a^{(L)})^2 \rangle $ 
and $\langle a^{(RS)} (a^{(P)})^2\rangle $.
However, as shown in Appendix~\ref{appRSPP}, the potentially very large terms $\langle a^{(RS)} (a^{(P)})^2 \rangle$ 
are actually exactly zero (because of the absence of correlations between RS and P).
Moreover, since $a^{(L)} \ll a^{(P)}$, we can neglect the $\langle a^{(RS)}(a^{(L)})^2 \rangle$ terms with respect 
to the $\langle a^{(P)} a^{(L)} a^{(RS)} \rangle$ terms.

Summarizing, we are left with two types of contributions to $\langle B_{\ell_1 \ell_2 \ell_3}\rangle$:
\beq
\langle B^{(RS)}_{\ell_1 \ell_2 \ell_3}\rangle = \sum_{m_1,m_2,m_3} 
\left( \begin{array}{ccc} \ell_1 & \ell_2 & \ell_3 \\ m_1 & m_2 & m_3  \end{array} \right)
\langle a^{(RS)}_{\ell_1 m_1} a^{(RS)}_{\ell_2 m_2} a^{(RS)}_{\ell_3 m_3} \rangle \, ,
\eeq
\beq
\langle B^{(PLRS)}_{\ell_1  \ell_2 \ell_3}\rangle
=  \sum_{m_1,m_2,m_3}  \left( \begin{matrix}  \ell_1 & \ell_2 & \ell_3 \cr m_1 & m_2 & m_3 \end{matrix} \right)
\langle a^{(P)}_{\ell_1 m_1} a^{(L)}_{l_2 m_2} a^{(RS)}_{l_3 m_3} \rangle + (5 \, \, {\rm permutations})   \, .
\eeq
We compute the first one in the following subsection, while the second in ~\cite{LENS}.

%%%%%%%%%%%%%%%%%%%%%%%%%%%%%%%%%%%%%%%%%%%

\subsection{Non-gaussianity from RS effect}

To compute the $\langle B_{\ell_1 \ell_2 \ell_3}^{(RS)}\rangle $ term is very simple. The only 
non-zero $a_{\ell m}^{(RS)}$ coefficients are those with $m=0$, so that:
\beq
 B_{\ell_1 \ell_2 \ell_3}^{(RS)} = 
\left( \begin{array}{ccc}
\ell_1 & \ell_2 & \ell_3 \\
0 & 0 & 0 
 \end{array} \right)~ a^{RS}_{\ell_1 0}~ a^{RS}_{\ell_2 0}~ a^{RS}_{\ell_3 0} \, .
\label{BlllRS}
\eeq
It is customary to define a reduced bispectrum $b_{\ell_1 \ell_2 \ell_3}$ via the following:
\beq
B_{\ell_1 \ell_2 \ell_3} = \sqrt{\frac{(2 \ell_1 +1)(2 \ell_2 +1)(2 \ell_3 +1)}{4 \pi}} 
%\left( \matrix{l_1 & l_2 & l_3 \cr 0 & 0 & 0} \right)
\left( \begin{array}{ccc} \ell_1 & \ell_2 & \ell_3 \\ 0 & 0 & 0  \end{array} \right)~ b_{\ell_1 \ell_2 \ell_3} \, .
\label{reduced}
\eeq

\begin{figure}[t!]  \begin{center}\begin{tabular}{c}
\includegraphics[width=8cm]{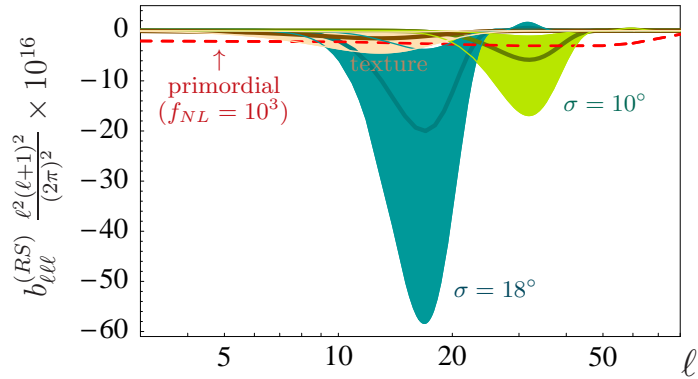} 
\put(-250,30){\large \rotatebox{90}{$ b^{(RS)}_{\ell \ell \ell}~ \frac{\ell^2 (\ell+1)^2}{(2 \pi)^2} \times 10^{16}$}} 
\put(-0,0){\Large $\ell$}
\put(-175,117){\BRed $\uparrow$} \put(-192,107){\BRed primordial } 
\put(-199,97){ ($f_{NL}=10^3$)\Black}
\put(-85,30){\Blue $\sigma=18^\circ$ \Black}
\put(-44,100){\Green $\sigma=10^\circ$ \Black}
\put(-124,115){\Orange \small texture \Black}
\end{tabular}\end{center}\vspace*{-0.5cm} 
\caption{Plot of $10^{16} \ell^2 (\ell+1)^2/(2 \pi)^2 ~ b^{(RS)}_{\ell \ell \ell}$ 
as a function of the multipole $\ell$, for $\sigma=18^\circ$ and $\sigma=10^\circ$, with 
$A = (7 \pm 3)\times 10^{-5}$. For comparison, we also plot in red the prediction for the
primordial signal for $f_{NL} =10^3$. 
The light (orange) shadow shows, for comparison, the result for the texture considered in~\cite{texture}.}
\label{fig-blllRS}\vskip 1cm
\end{figure}

We plot the diagonal contribution $b^{(RS)}_{\ell \ell \ell}$ in fig.~\ref{fig-blllRS}: the equivalent amplitude is very high compared 
to a typical primordial signal, since it corresponds roughly to $f_{NL}\sim 10^4$. However, for $\ell \gtrsim 60$ the 
RS signal due to the large Void goes rapidly to zero.  
Therefore, for experiments like WMAP or Planck, which go up to about $\ell \sim 800$ and $\ell \sim 2000$ respectively,
only a small subset of the data is affected by the RS contribution.

Focusing on the RS signal, we now turn to estimate whether it is detectable or not. The signal on a single multipole is lower than the 
cosmic variance: so we have to sum over all the $\ell$'s to find a bispectrum Signal-to-Noise ($S/N$) ratio.
For a signal labeled by $i$, this is defined as (see {\it e.g.}~\cite{review}):  
\begin{equation}
(S/N)_i=\frac{1}{\sqrt{F_{ii}^{-1}}}  ~~ , ~~~~~
F_{ii}= \sum_{2 \leq l_1 \leq l_2 \leq l_3 \leq l_{\rm max}} \frac{ (B^{(i)}_{l_1 l_2 l_3})^2}{\sigma^2_{\ell_1 \ell_2 \ell_3}}   \, ,
\label{Fiidef}
\end{equation}
where $\sigma_{\ell_1 \ell_2 \ell_3}$ is the variance of the bispectrum: 
\beq
\sigma^2_{\ell_1 \ell_2 \ell_3}\sim \langle {\cal C}_{\ell_1} \rangle \langle {\cal C}_{\ell_2}\rangle \langle {\cal C}_{\ell_3}
\rangle \Delta_{\ell_1 \ell_2 \ell_3} \, ,
\eeq 
and $\Delta_{\ell_1 \ell_2 \ell_3}=1,2,$ or $6$ respectively if all $\ell$'s are different, if only two of them are equal or 
if they are all equal. 
The ${\cal C}_\ell$'s represent the sum of the CMB power spectrum plus the power spectrum of the noise of the detector. 
In general, at some $\ell_{max}$ the noise becomes dominant, while below $\ell_{max}$ the variance is dominated by the Primordial one, namely
${\cal C}_\ell \simeq \langle C_\ell^{(P)} \rangle$. 

\begin{figure}[t!] \begin{center}\begin{tabular}{c}
\includegraphics[width=7.8 cm]{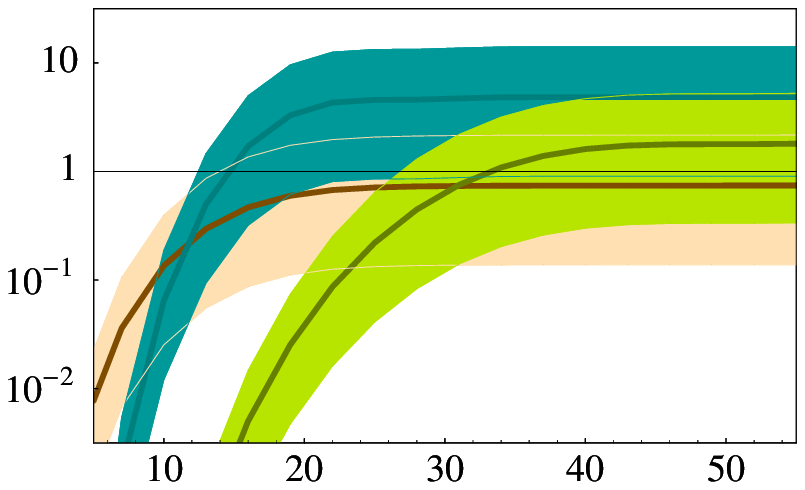} 
\put(-270,80){\Large  $ {(S/N})_{RS}$} \put(-0,0){\Large $\ell_{max}$}
\put(-170,122){\Blue $\sigma=18^\circ$ \Black}
\put(-124,35){\Green $\sigma=10^\circ$ \Black}
\put(-54,55){\Orange texture \Black}
~~~~~~~~~~~~~~~~~~~~~~~~~\includegraphics[width=5cm]{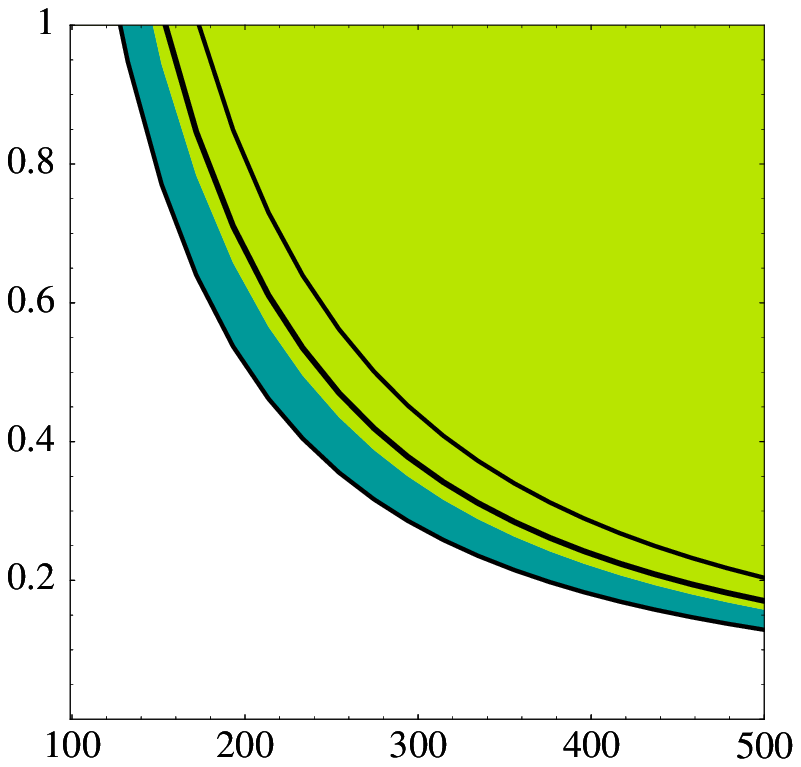} 
\put(-86,-8){  $L ~[{\rm Mpc}/h]$ } \put(-163,80){\Large $|\delta_0|$}
\put(-85,105){\Black \rotatebox{0}{$(S/N)_{RS}>1$} \Black}
\put(-85,90){\Black for $\sigma \ge 10^\circ$ \Black}
\put(-125,45){\Black for $\sigma \ge 18^\circ \nearrow$ \Black}
\end{tabular}\end{center}\vspace*{-0.5cm} 
\caption{Left: Plot of $S/N$ for the RS as a function of the multipole $\ell_{max}$ for
$\sigma=18^\circ$ and $\sigma=10^\circ$, with $A = (7 \pm 3)\times 10^{-5}$.  
The light (orange) shadow shows, for comparison, the analogous quantity for the texture considered in~\cite{texture}.
Right: The black lines, from bottom to top, show the contour levels for $A= (4,7, 10) \times 10^{-5}$ in the plane $( L,|\delta_0|)$.
For $\sigma=10^\circ$, all the points of the light shaded (green) region would give $(S/N)_{RS} >1$. 
For $\sigma=18^\circ$ this region gets larger and includes the dark shaded one, which goes down to $A=4\times 10^{-5}$.}
\label{SNRS}\vskip 1cm
\end{figure}

Neglecting the instrumental noise, in the left panel of fig.~\ref{SNRS} we show the result for $(S/N)_{RS}$ as a function of $\ell_{max}$.
As one can see, the signal is detectable for a large part of the parameter space: so it should already be possible to look for such a signal 
in the WMAP data. 
Conversely, the absence of any detection would give interesting constraints on the size $L$ and density $\delta_0$ of a large Void. 
We show in the right panel of fig.~\ref{SNRS} the region of physical parameter space which would give rise to a detectable signal, namely 
$(S/N)_{RS} >1$: for a cold region diameter $\sigma \ge 10^\circ$, this happens for all the points in the light shaded region.
For $\sigma\ge18^\circ$ such region gets larger and includes the dark shaded one. For instance, for a Void with $\sigma=18^\circ$ 
and $\delta_0=-50\%$, a signal in the bispectrum would appear only if $L\ge 200 {\rm Mpc}/h$, which corresponds to $A\ge 4\times 10^{-5}$. 

Other candidates for a structure that could explain the Cold Spot might be subject to constraints analogous to those discussed above 
for a big Void. For instance, from the left panel of fig.~\ref{SNRS} one can see that the texture proposed in~\cite{texture} 
would give rise to $S/N>1$ in the bispectrum only if $A\gtrsim 7\times 10^{-5}$.

%%%%%%%%%%%%%%%%%%%%%%%%%%%%%%%%%%%%%%%%%%%%%%%%%%%%%%%%%%%%%%%%%%%%%%%%%%%%%%%%%%%%%%

\subsection{Contamination of $f_{NL}$ measurements}
\label{deltafNLSECT}

We now turn to the impact that a huge Void in the line of sight would have on the measurement of the primordial non-gaussianity 
parameter $f_{NL}$, in terms of which it is customary to parametrize a primordial non-Gaussian signal.
By definition, $f_{NL}$ is introduced (see {\it e.g.}~\cite{review} for details) 
parameterizing the primordial curvature perturbations $\phi(x)$ as follows:
\beq
\phi(x)=\phi_L(x) + f_{NL} (\phi^2_L(x)-\langle\phi^2_L(x) \rangle)
\eeq
where $\phi_L(x)$ is the linear Gaussian part of the perturbation.
Given a physical model (like slow-roll inflation)  $f_{NL}$ is generically a function of the momenta, {\it i.e.} $f_{NL}(k)$,
but in the quantitative data analyses it is usually assumed to be just a constant number.
Note that single field minimally coupled slow-roll inflationary models predict very small values for $f_{NL}$, 
that is $f_{NL} = {\cal O}(0.1)$~\cite{Maldacena,Acquaviva}, but other models may predict larger values (see {\it e.g.}~\cite{review}).
The primordial bispectrum coefficients can be written as:
\beq
B^{prim}_{l_1 l_2 l_3}= f_{NL} \, \tilde{B}^{prim}_{l_1 l_2 l_3} \, ,
\eeq
where the $ \tilde{B}^{prim}_{l_1 l_2 l_3}$ have a specific form in terms of the primordial spectrum 
and the radiation transfer function.

As we have seen, the RS effect leads to a large contribution to $\langle B_{\ell_1 \ell_2 \ell_3} \rangle $ for multipoles in the range 
$10\leq \ell \leq 50$. 
On the contrary, the Lensing effect~\cite{LENS} is much smaller at low $\ell$'s, but could contaminate the primordial bispectrum 
signal at large $\ell$'s, 
since it couples the low RS-$\ell$'s with the high $\ell$'s of the primordial signal~\cite{LENS}.
The impact on $f_{NL}$ due to the RS effect can be computed by estimating the following ratio~\cite{review,cooray}:
\beq
\Delta f^{(RS)}_{NL}(\ell_{max})=
\frac{\sum_{2 \le \ell_1 \le \ell_2 \le \ell_3 \leq \ell_{max}}
        \frac{B^{(RS)}_{\ell_1 \ell_2 \ell_3} \tilde{B}^{prim}_{\ell_1 \ell_2 \ell_3}}{\sigma^2_{\ell_1 \ell_2 \ell_3}} }
     {\sum_{2 \le \ell_1 \le \ell_2 \le \ell_3 \leq \ell_{max}}
        \frac{(\tilde{B}^{prim}_{\ell_1 \ell_2 \ell_3})^2}{\sigma^2_{\ell_1 \ell_2 \ell_3}}} ~~~ .
\label{deltafNLdef}
\eeq
In other words, if a large Void exists, one should subtract it from the data in order to get the correct value for $f_{NL}$, 
thus avoiding to overestimate the latter by the amount $\Delta f^{(RS)}_{NL}$.

We may very easily give an approximation of the $\tilde{B}^{prim}_{\ell_1 \ell_2 \ell_3}$ appearing in the numerator 
of~(\ref{deltafNLdef}) by using~(\ref{reduced}) with the Sachs-Wolfe approximation for the primordial signal: 
\beq
\tilde{b}^{prim}_{\ell_1 \ell_2 \ell_3}=-6 (\langle C^{(P)}_{\ell_1} \rangle \langle C^{(P)}_{\ell_2} \rangle +
                                            \langle C^{(P)}_{\ell_2} \rangle \langle C^{(P)}_{\ell_3} \rangle +
                                            \langle C^{(P)}_{\ell_1} \rangle \langle C^{(P)}_{\ell_3} \rangle) ~~ .
\eeq
This is a good approximation only for the low-$\ell$ plateau in the power spectrum, while the full expression should be used for higher $\ell$'s. 
However, since $\langle B^{(RS)}_{\ell_1 \ell_2 \ell_3}\rangle$ vanishes if any of the $\ell$'s is larger than about $60$, 
we consider it to be a sufficiently fair approximation to get an estimate of the numerator of~(\ref{deltafNLdef}). 
This fact also allows to neglect the experimental noise in the numerator's $\sigma^2_{\ell_1 \ell_2 \ell_3}$. 
Instead the denominator is a quantity very sensitive to the experiment (see~\cite{review}). 
In fact, at some high $\ell_{max}$ (dependent on the experiment) the experimental noise in the $\sigma^2_{\ell_1 \ell_2 \ell_3}$ 
of the denominator becomes so large that the multipoles $\ell> \ell_{max}$ give a negligible contribution to the sum.
Accordingly, the denominator of~(\ref{deltafNLdef}) turns out to be equal to $5.8 \times 10^{-2}$ for WMAP ($l_{max}\approx 800$) and 
$0.19$ for Planck ($l_{max}\approx 2000$)~\cite{review}, so that one obtains:
\beq 
\Delta f^{(RS)}_{NL} \approx \begin{cases}  1 & {\rm for \, WMAP} \\
                                         0.1  & {\rm for  \, Planck} \end{cases}~~~.
\eeq

\begin{figure}[t!]  \begin{center}\begin{tabular}{c}
\includegraphics[width=7. cm]{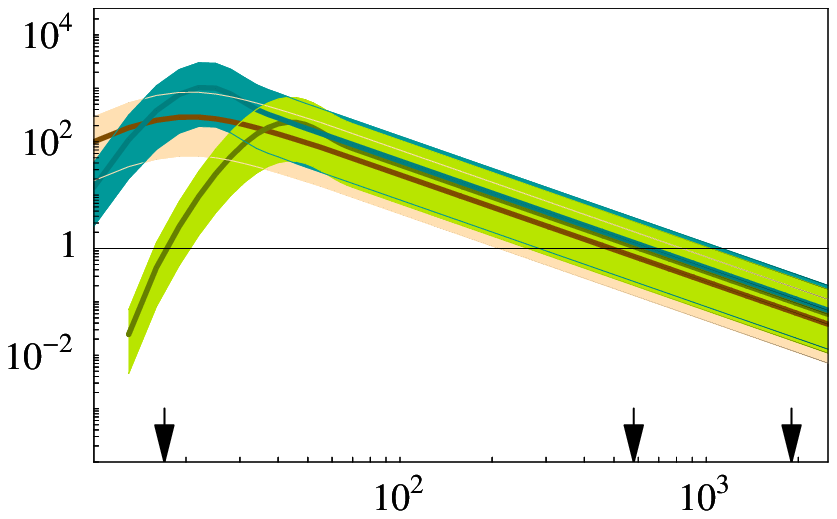} 
\put(-215,45){\rotatebox{90}{\large  $ \Delta f^{(RS)}_{NL}(\ell_{max})$}} \put(-0,0){\Large $\ell_{max}$}
\put(-170,112){\Blue $\sigma=18^\circ$ \Black}
\put(-152,56){\Green $\sigma=10^\circ$ \Black}
\put(-160, 30){COBE} \put(-70, 30){WMAP} \put(-30, 30){Planck} 
$~~~~~~~~~~~~~~~~~~~$ \includegraphics[width=7.3 cm]{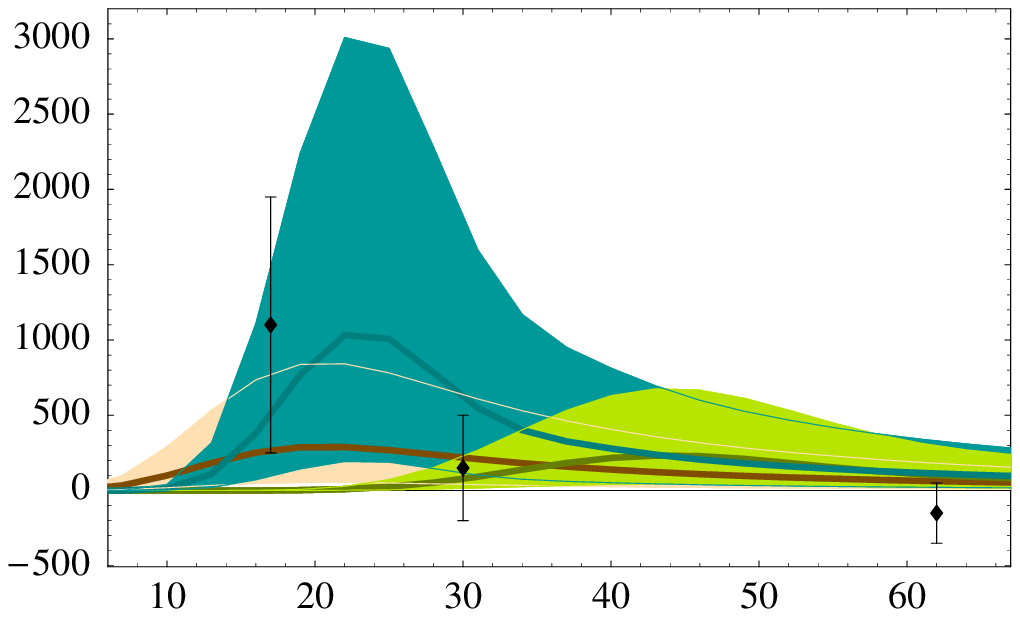} 
\put(-225,45){\rotatebox{90}{\large  $ \Delta f^{(RS)}_{NL}(\ell_{max})$}} \put(-0,0){\Large $\ell_{max}$}
\put(-110,48){\Huge $\swarrow$} \put(-85,64) {$f_{NL}$ at 1-$\sigma$} \put(-82,53){from WMAP1}
\end{tabular}\end{center}\vspace*{-0.5cm} 
\caption{The correction to $f_{NL}$ arising from the inclusion of the RS effect, 
$\Delta f^{(RS)}_{NL} (\ell_{max})$, plotted  as a function of $\ell_{max}$, for 
$\sigma=18^\circ$ and $\sigma=10^\circ$, together with $A = (7 \pm 3)\times 10^{-5}$.
Left: the log-log scale allows a better understanding of the effect for different experimental sensitivities.
Right: the linear scale is more useful for a comparison with the constraints on $f_{NL}$ reported by the WMAP 1-year 
analysis~\cite{WMAP1GAUSS}. The light (orange) region shows the analogous results in the texture case~\cite{texture}.}
\label{deltafNL} \vskip .5cm
\end{figure}

It is also interesting to study the dependency of $\Delta f^{(RS)}_{NL}(\ell_{max})$ for smaller values of $\ell_{max}$. 
This can be easily done because the denominator in~(\ref{deltafNLdef}) is well approximated by $\ell_{\rm max} \times 10^{-4}$~\cite{Komatsu:2001rj},
until the experimental noise is negligible. The result is plotted in fig.~\ref{deltafNL}:
from the left panel it turns out that the RS effect due to the large Void does not affect much the $f_{NL}$ measurements 
for high resolution experiments. 
In fact, the corrections are localized at low $\ell$'s because $\langle B_{\ell_1 \ell_2 \ell_3} \rangle$ get a sizable RS correction 
only if $\ell_1, \ell_2, \ell_3 \lesssim 60$, while the search for a primordial non-gaussianity uses all the experimental 
data-points up to $\ell_{\max}\sim 800-2000$. 
Therefore, even if the multipoles $\ell\lesssim 60$ are strongly affected, they only represent a small 
fraction of all the bispectrum data-points. 
Note also that the impact on $f_{NL}$ is roughly the same for $\sigma=18^\circ$ 
and $\sigma=10^\circ$: the reason is indeed that a smaller Void affects higher $\ell$'s, 
which are more relevant for the extraction of the primordial $f_{NL}$.   

The right plot in fig.~\ref{deltafNL} allows a direct comparison with the WMAP-1year experimental constraints on $f_{NL}$~\cite{WMAP1GAUSS}. 
We can already see that some region of the parameter space can be excluded for the Void with $\sigma=18^\circ$: 
values for the amplitude $A \ge 7(8) \times 10^{-5}$ are excluded at $1(2)$-$\sigma$ by the error bar localized at $\ell=30$. 
Of course, such an analysis would require a more refined technical treatment -- for example including the recent 
WMAP 5-year data, sky-cuts, the full expression for $B^{prim}_{\ell_1 \ell_2 \ell_3}$ -- which we do not address in the present paper. 
A full analysis should also include the correlation matrix with other cosmological sources: SZ-lensing effect, point sources 
and the primordial signal. 
Anyway, looking for $f_{NL}$ is not the best way to constrain a large Void: one should rather compare directly the observed 
bispectrum data with the prediction for $\langle B^{(RS)}_{\ell_1 \ell_2 \ell_3}\rangle$.
Since we have shown in fig.~\ref{SNRS} that the RS Signal-to-Noise is above unity 
for most of the parameter space, a full analysis should be able to set interesting constraints. 

Finally, we may also give a comment on the results by~\cite{Yadav}, whose authors claim a detection of primordial non-gaussianity 
already in the WMAP data. In their analysis they also find that, masking the region which corresponds to the Cold Spot, 
their measured $f_{NL}$ goes up by an amount of $7$. In the light of our analysis, this cannot be due to the RS effect, 
because the RS contribution to $f_{NL}$ is smaller (about $1$) for WMAP and, moreover, has opposite sign.
The possibility that this could be ascribed to the Lensing effect will be discussed in~\cite{LENS}.

%%%%%%%%%%%%%%%%%%%%%%%%%%%%%%%%%%%%%%%%%%%%%%%%%%%%%%%%%%%%%%%%%%%%%%%%%%%%%%%%%%

\section{Other Voids}
\label{multivoids}

We briefly comment in this section about the possibility that several Void regions are present in the sky. 
%It has been pointed out by~\cite{manyvoids,Szapudi} that the detection of the ISW signal (via correlation between the 
%galaxy surveys and the CMB) seems to be due to several regions in the sky (about 50 in~\cite{Szapudi}, all located in 
%a region that covers about $10\%$ of the sky)  of radius of about $5^\circ$, and amplitude $A \approx 4\times 10^{-6}$, on average.
The authors of ~\cite{manyvoids,Szapudi} have claimed detection of the ISW effect because it could explain the correlations  
that they were able to identify between the galaxy surveys and the CMB data. 
By exploring a region of the northern hemisphere that covers about $20\%$ of the sky, 
\cite{Szapudi} has catalogued about 50 of such Voids\footnote{The Cold Spot, located in the southern hemisphere, is clearly 
not among them.}, 
which have a mean radius of about $5^\circ$ and 
a mean amplitude $\bar A \approx 3.6 \times 10^{-6}$ (or, equivalently, $\Delta T \approx-11\mu$K);
their density contrast for the luminous matter can be directly observed in the galaxy surveys;
the knowledge of their redshift %(detected in the galaxy surveys)  
also allows to estimate their physical size $L$, which turns out to be of about $50-100 {\rm Mpc}/h$. 
As stressed in the introduction, \cite{SarkarVoids} 
has interestingly pointed out that the existence of such large Voids is unlikely to be explained by the standard 
structure formation scenario. It could be relevant, therefore, to apply our considerations to these Voids as well. 

If $N$ Voids are present in the sky, however, there are differences with respect to the analysis developed in the previous sections.
If we focus on the two-point correlation function, in addition to the $N$ terms of the same type as~(\ref{ClRS}),
we have to compute also $N^2-N$ interference terms. 
As for the three-point correlation function, it would contain $N$ terms of the same kind as~(\ref{BlllRS}), 
plus $N^3-N$ interference terms. 
For a random distribution of Voids in the sky, one expects the interference terms to add up incoherently,
leading only to an oscillatory modulation of the correlation functions.

For a precise computation of the effect of these Voids on the power spectrum and bispectrum, 
it would be necessary to know, for each Void, its location, 
the amplitude $A$ and the angular size $\sigma$ of its temperature profile.
Note that the most of the contribution is expected to come from the few Voids with the largest $A$ and $\sigma$.
The authors of~\cite{Szapudi} give these informations for the 50 Voids identified, except for the amplitude $A$, 
for which only the mean value $\bar A $ is provided. 

\begin{figure}[t!] \begin{center}\begin{tabular}{c}
\includegraphics[width=8cm]{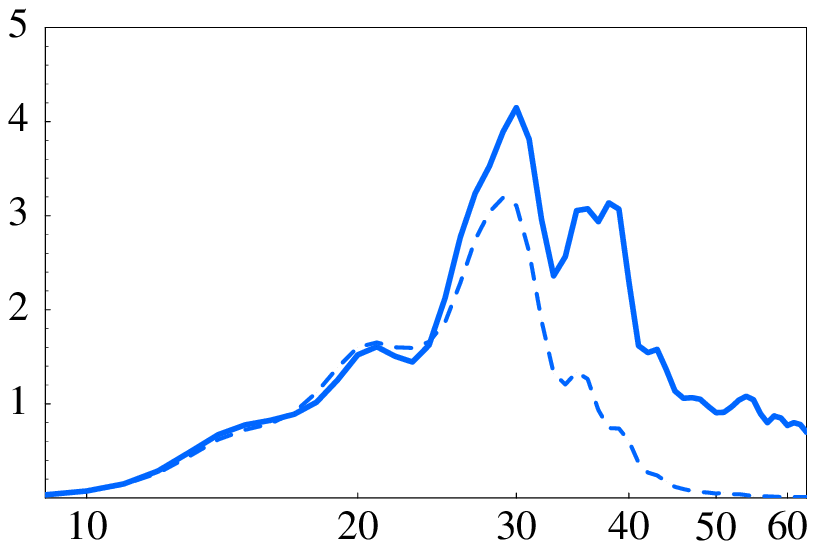} 
\put(-235,35){\rotatebox{90}{$ C^{(RS)}_{\ell} \frac{\ell (\ell+1)}{2 \pi} T_0^2 [\mu K^2]$}} \put(-0,0){\Large $\ell$}
%$\put(-65,94){\Blue $\sigma=18^\circ$ \Black} \put(-37,38){\Green $\sigma=10^\circ$ \Black}
%$\put(-149,25){\Orange \rotatebox{15}{\small texture} \Black}
~~~~~~~~~~~~~~~\includegraphics[width=5cm]{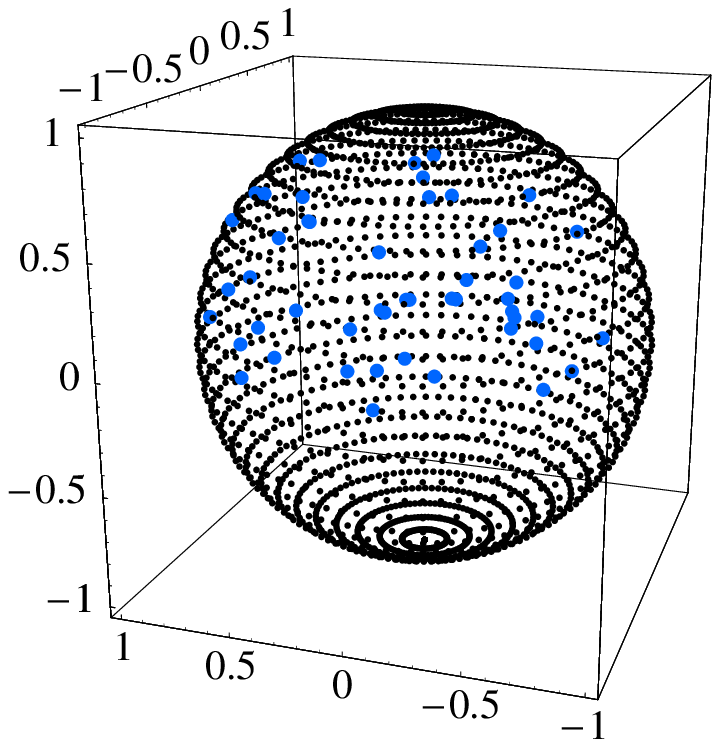} 
%\put(-215,40){\rotatebox{90}{$ \langle C_{\ell}\rangle \frac{\ell (\ell+1)}{2 \pi} T_0^2 [\mu K^2]$}} \put(-0,0){\Large $\ell$}
\end{tabular}\end{center}\vspace*{-0.5cm} 
\caption{In the left panel we plot (solid line) the RS power spectrum coefficients 
obtained from the 50 Voids identified in~\cite{Szapudi}, 
for all of which we assigned the mean amplitude $A=3.6 \times 10^{-6}$. These Voids have $\sigma$ in the range $3^\circ-14^\circ$: 
the dashed curve is obtained by selecting (among the 50 Voids) the 19 ones having $\sigma \gtrsim 10^\circ$.
The dotted thin curve is the full result without interference terms.
Right: a sketch of the location of the 50 Voids~\cite{Szapudi}.}
\label{fig-multipleVoids}\vskip 1cm
\end{figure}

We can nevertheless give an estimate of the impact on the CMB due to the RS effect of the Voids catalogued in~\cite{Szapudi}
(their location is sketched in the right panel of fig.~\ref{fig-multipleVoids}),
by assigning to each of them an amplitude equal to the mean one. The angular size of the Voids spans from $3^\circ$ to $14^\circ$,
and there are 19 Voids with $\sigma\gtrsim 10^\circ$.
The left panel of fig.~\ref{fig-multipleVoids} shows the result for the two-point correlation function
(where we also show that the interference terms are negligible), 
whose order of magnitude can be understood by the following easy argument.
Rescaling to the case of an amplitude $\bar A$ 
the results obtained in sect.~\ref{powerspectrum} for one Void with radius of about $5^\circ$,
one gets: $C^{(RS)}_{\ell} \frac{\ell (\ell+1)}{2 \pi} T_0^2\approx 0.1 \mu {\rm K}^2$.
Assuming that the interference terms (which depend on the relative location in the sky of the Voids) are irrelevant,
the latter value for a single Void has to be multiplied by $N=50$, thus obtaining 
$C^{(RS)}_{\ell} \frac{\ell (\ell+1)}{2 \pi} T_0^2 ={\cal O} (5 \mu {\rm K}^2)$, 
consistently with the left plot of fig.~\ref{fig-multipleVoids}. 
%An upper bound can be given in the case in which all the objects maximally interfere. Note that obviously this is largely 
%overestimating the effect, since this case would correspond to the configuration in which all objects are all aligned and located 
%at the same position in the sky.
%Nonetheless in this case the two point correlation function will give an effect of the order of $N^2$ times the effect for a single 
%patch, that is $250 \mu K^2$.
We may estimate the contribution to the three point function similarly.
The signal-to-noise ratio for a Void of radius $5^\circ$ and amplitude $\bar A$ would be of about $10^{-3}$, as shown in 
sect.~\ref{bispectrum}. As done before, the final result is obtained by multiplying by the number of Voids, 
leading to $(S/N)_{RS} = {\cal O} (5 \times 10^{-2})$, assuming again that the interference terms are negligible.

%The upper bound instead would be given by multiplying by $50^3$, leading to $S/N \approx 10^2$.
%We did not compute exactly this effect, but one could expect it to be closer to the lower than to the upper bound 
%(as it happens for the two-point correlation function).

%%%%%%%%%%%%%%%%%%%%%%%%%%%%%%%%%%%%%%%%%%%%%%%%%%%%%%%%%%%%%%%%%%%%%%%%%%%%%%%

\section{Conclusions}
\label{concl}

Motivated by the so-called Cold Spot in the WMAP data,
we have studied in this paper the impact on statistical CMB predictions, in particular the two and three point correlation functions, 
of the presence of an anomalously large Void along the line of sight. 
Indeed, the existence of such a Void could be at the origin of the Cold Spot, due to the Rees-Sciama effect 
(the Lensing effect is analyzed in the companion paper~\cite{LENS}). 

First, we have computed the temperature profile using an LTB solution of the Einstein equations, matched to an FLRW metric.
Then, we have computed its impact on statistical predictions for the CMB, {\it assuming} this structure 
to be uncorrelated with the Primordial fluctuations. 
As suggested by~\cite{texture}, we consider the angular size of such a Void to be about $10^\circ-18^\circ$ and 
its temperature at the centre to be characterized by $\Delta T=-(190\pm 80) \mu K$.

For the  power spectrum the results are as follows.
The RS effect is non-negligible: %using the amplitude and size suggested by~\cite{texture} for the temperature profile, 
we predict a bump of $5 \% -25\%$ to be added to the Primordial spectrum, localized at multipoles $5 \leq \ell \leq 50$.
This should lead to a variation in the $\chi^2$ for the WMAP fits of order unity.

Then we have studied the impact on the bispectrum coefficients. For the RS effect we have found that the Signal-to-Noise 
ratio is larger than unity at $\ell \gtrsim 40$ for most of the parameter space, and therefore this should already be visible 
in the available data.
Through the bispectrum, we have studied also the impact of such a structure on the determination of the primordial non-gaussianity. 
The RS bispectrum signal is large but localized at low multipoles ($10 \leq \ell \leq 50 $), so it has a small impact on 
high resolution 
experiments, which can go up to very large multipoles: the overestimation of $f_{NL}$ due to the RS effect turns out to be 
$\Delta f_{NL}^{(RS)}\simeq 1$ for WMAP and $\Delta f_{NL}^{(RS)}\simeq 0.1$ for Planck.
Using the already existing WMAP1-year constraints~\cite{WMAP1GAUSS} on $f_{NL}$ at low $\ell$, one can exclude extreme values of the 
temperature contrast of the Void. For example, values for $\Delta T/T$ larger than $8\times 10^{-5}$ for a Cold Spot with diameter 
of $18^\circ$ are likely to be excluded, via a full analysis. 
So, we conclude that the bispectrum is a valuable tool for constraining an anomalously large Void, through the RS effect.

We have also considered the 50 Voids whose detection has been claimed in~\cite{Szapudi}. 
These Voids have mean angular diameter of about $10^\circ$ and average temperature 
at the centre characterized by $\Delta T \approx -11 \mu K$. 
The effect on the two-point correlation function is about $0.5\%$, hence much smaller than the one due to the 
large Void considered to explain the Cold Spot. 
The effect on the three-point correlation function has not been studied in detail 
but it is expected to lead to a Signal-to-Noise ratio smaller than one.

Finally, we have also applied our considerations to the case in which the Cold Spot is explained by a 
cosmic texture~\cite{texture} rather than a large Void: the effect on the power spectrum is similar but somewhat 
smaller; the three-point correlation function leads also to a smaller Signal-to-Noise ratio, 
which however turns out to be above unity for some part of the parameter space.

\acknowledgments We would like to thank Tirtho Biswas, Marcos Cruz, Paul Hunt, Bob McElrath, Oystein Rudjord and Subir Sarkar 
for useful discussions and suggestions.

%%%%%%%%%%%%%%%%%%%%%%%%%%%%%%%%%%%%%%%%%%%%%%%%%%%%%%%%%%%%%%%%%%%%%%%%%%%%%%%%%%

\appendix

\section{Vanishing of the P-P-RS contribution to the Bispectrum}
\label{appRSPP}

We show here that the term containing $\langle a^{(P)} a^{(P)} a^{(RS)} \rangle$ in the bispectrum 
coefficients $\langle B_{\ell_1 \ell_2 \ell_3} \rangle$ is vanishing. 
More generally, any term of the kind $\langle a^{(P)} a^{(P)} a^{(i)} \rangle $ is zero, for any $a^{(i)}$ 
uncorrelated with $a^{(P)}$.

Averaging~(\ref{defBlm}) over a statistical ensemble of possible realisation for the Universe one gets:
\bea
\langle B^{m_1 m_2 m_3}_{~\ell_1~ \ell_2~ \ell_3} \rangle^{(PPRS)} & =& 
  \delta_{m_1 0} \langle  a_{\ell_1 0}^{RS} ~ a_{\ell_2 m_2}^P ~ a_{\ell_3 m_3}^P \rangle 
+ (1,2,3\rightarrow 2,3,1) + (1,2,3\rightarrow 3,1,2) ~~\\
&=&  \delta_{m_1 0~} a_{\ell_1 0}^{RS} (-)^{m_3} \langle a_{\ell_2 m_2}^P ~ {a^*}_{\ell_3, -m_3}^P \rangle 
      +  (1,2,3\rightarrow 2,3,1) + (1,2,3\rightarrow 3,1,2)~~ \nonumber \\
&=&
\delta_{m_1 0~} a_{\ell_1 0}^{RS}(-)^{m_3} C_{\ell_2} \delta _{\ell_2 \ell_3} \delta_{m_2, -m_3} 
+ (1,2,3\rightarrow 2,3,1) + (1,2,3\rightarrow 3,1,2)
~~. \nonumber
\eea
Summing over the $m_i$'s according to~(\ref{defBl}), one finds 
\beq
\langle B_{\ell_1 \ell_2 \ell_3} \rangle ^{(PPRS)}= 
a_{\ell_1 0}^{RS} \langle C_{\ell_2} \rangle \delta_{\ell_2, \ell_3} 
\underbrace{ \sum_{m} (-)^{m_2} \left( \begin{array}{ccc} \ell_1 & \ell_2 & \ell_2 \\ 0 & m_2 & -m_2  
\end{array} \right)}_{=\delta_{\ell_1 0}}
+ (1,2,3\rightarrow 2,3,1) + (1,2,3\rightarrow 3,1,2) =0 ~~,
\eeq
where we have used a known property of the Wigner $3-j$ symbol and the fact that by definition $a_{00}^{RS}=0$, see~(\ref{defT}).

\end{document}